\documentclass[pre,twocolumn]{revtex4}%
\usepackage{amsfonts}
\usepackage{amsmath}
\usepackage{amssymb}
\usepackage{graphicx}
\usepackage{aas_macros}%

\begin{document}

\title{Pauli blocking effects in thermalization of relativistic plasma}
\author{M. A. Prakapenia}
\affiliation{ICRANet-Minsk, B. I. Stepanov Institute of Physics, National Academy of Sciences of Belarus, 220072 Nezale\v znasci Av. 68-2, Minsk, Belarus}
\affiliation{Department of Theoretical Physics and Astrophysics, Belarusian State University, 220030 Nezale\v znasci Av.  4, Minsk, Belarus}
\author{G. V. Vereshchagin}
\affiliation{ICRANet, 65122 Piazza della Repubblica, 10, Pescara, Italy}
\affiliation{INAF – Istituto di Astrofisica e Planetologia Spaziali, 00133 Via del Fosso del Cavaliere,100, Rome, Italy}

\begin{abstract}
We investigate the effects of Pauli blocking on thermalization process of relativistic plasma by solving relativistic Uehling-Uhlenbeck equations with QED collision integral for all binary and triple processes. With this purpose we consider nonequilibrium initial state of plasma to be strongly degenerate. We found that when electron-positron annihilation is active, initial plasma degeneracy is quickly destroyed. As a result in a wide range of final temperatures ranging from nonrelativistic to mildly relativistic $0.1 m_e c^2 \leq k_B T\leq 10 m_e c^2$ thermalization is not affected by Pauli blocking. Conversely, when electron-positron annihilation process is inactive, thermalization process in such degenerate plasma is strongly affected by Pauli blocking. This is possible either in a nonrelativistic plasma, with equilibrium temperature $k_B T\leq 0.3 m_e c^2$, or in photon-electron plasma. In these cases all reaction rates are strongly suppressed by Pauli blocking and thermalization does not occur until electrons can populate energy states above the Fermi energy. Soon after this happens thermalization proceeds suddenly in an avalanche-like process.
\end{abstract} 

\maketitle

\section{Introduction}

Pauli exclusion principle is a fundamental principle of quantum mechanics and
it manifests in many branches of physics: condensed matter, chemistry,
molecular biology, etc. Such phenomena as insulator-conductor transition or
Mott effect, Feshbach resonance and atom-exchange chemical reactions
\cite{ebeling2009,bruun2009,wille2008,ospelkaus2010,demarco2019} deal with
Pauli blocking. Pauli principle plays a crucial role in a dense matter state
\cite{ebelingbook}, it affects conductivity in a dense Coulomb plasma
\cite{shmidt1989}, for extreme plasma densities it leads to depression of
ionization potential \cite{roepke2019}, it is crucial in many nuclear physics
problems \cite{xing2017,ray1993,bertulani2010,stefanini2019}. The development
of new techniques for atomic gases cooling allows to create Fermi gases in a
laboratory. Suppression of interactions in such a degenerate Fermi gases is
subject of active research
\cite{demarco1999,demarco2001,ferrari1999,aikawa2014}. Lattice Fermi gas can
serve as direct probe of Pauli blocking \cite{omran2015}. Existence and
stability of such compact astrophysical objects as white dwarfs \cite{1931ApJ....74...81C,1935MNRAS..95..207C} and neutron
stars is possible due to Pauli blocking  \cite{2000csnp.conf.....G}, which
modifies the equation of state at high densities and prevents gravitational
collapse of stars with mass below $\sim3$ solar masses after their nuclear
fuel is exhausted.

Another arena of manifestation of the Pauli blocking effect is relativistic
plasma. In addition to astrophysical environments, it is also of interests in
laboratory experiments targeting observation of QED processes \cite{ruffini2010}.
Attempts to create electron-positron plasma with high power lasers or laser
interactions with matter \cite{brady2013,chen2014,dipiazza2012,sarry2015} are
under way. Such plasma is typically formed out of equilibrium. Thermalization
process of non-equilibrium relativistic plasmas has been studied
\cite{arv2007,arv2009}, and thermalization timescales were determined \cite{2010PhRvE..81d6401A}, but so far Pauli blocking in this process has been neglected.

Kinetic equations accounting for Pauli blocking and Bose enhancement effects
are called Uehling-Uhlenbeck equations. These equations correctly describe the
limit of fully degenerate state, when  Pauli blocking reduce reaction rates of
fermions to zero. The present work deals with optically thick relativistic
plasma composed of electron, positrons and photons \cite{bookveresh} and
focuses on strongly degenerate initial plasma states. The paper is organized
as follows. In Section II conditions of plasma degeneracy are reviewed. In
Section III relativistic Boltzmann equations are introduced. In Section IV
thermalization process of superdegenerate plasma is discussed. The main
results are summarized in Section V. In appendix we present a table describing Bose enhancement and Pauli blocking factors for all binary and triple reactions considered.

\section{Fermion critical density}

\begin{figure}[ptb]
\center\includegraphics[width=80mm]{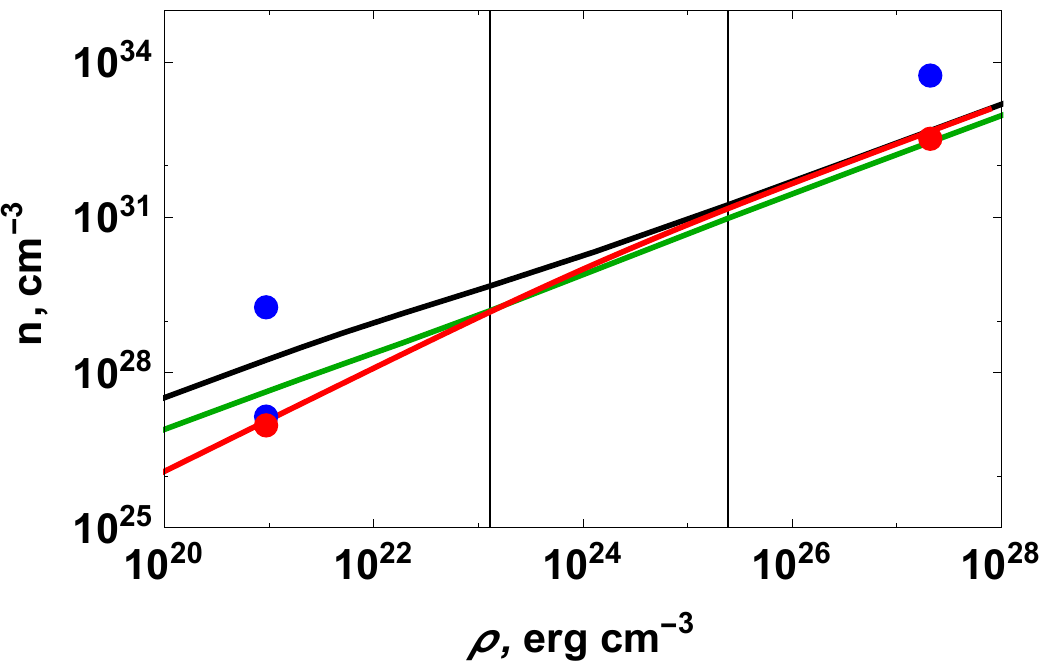}\caption{Number density - energy
density diagram of a photon-electron-positron plasma. Green curve corresponds
to thermal equilibrium state. Black curve shows the transition from
nondegenerate $D>1$ to degenerate $D<1$ plasma, where $D$ is defined by
Eq.~(\ref{D}). Red curve corresponds to fully degenerate pair state defined by
Eq.~(\ref{nmaxdef}). Vertical line on the left corresponds to the transition from
nonrelativistic to relativistic pair plasma ($\theta_{f\!in}=0.3$). Vertical line on the right corresponds to relativistic pair plasma with $\theta_{f\!in}=1$.
(Initial conditions for calculations are denoted by dots for pairs (red) and photons (blue).}%
\label{initcond}%
\end{figure}

The degree of plasma degeneracy is characterized by the parameter \cite{degrootbook}%
\begin{equation}
D=\frac{1}{n\lambda_{th}^{3}},\label{D}%
\end{equation}
where $n$ is number density of particles in plasma, $\lambda_{th}=\dfrac{c\hbar}{kT}$ is
the thermal wave-length, $k$ is Boltzmann constant, $T$ is temperature, $\hbar=h/(2\pi)$, $h$ is Planck constant. The number density -- energy density diagram for relativistic electron-positron plasma is presented in Fig.~\ref{initcond}. The black line corresponds to $D=1$; below this curve $D>1$ and plasma is nondegenerate, while above this curve $D<1$ and plasma is degenerate. Green curve corresponds to thermal equilibrium state. Red curve shows the maximum number density for electron-positron pairs (fully degenerate state). Note that thermal equilibrium state is very near to the border $D=1$.

Regarding fermions only (electrons and positrons without photons) it is important to note that both fully degenerate and thermal states have comparable number density of particles for ultrarelativistic average energy per particle (see
Fig. 2, bottom), specifically number density for fully
degenerate state is
\begin{equation}
n_{cr}=\frac{8\pi}{3 h^{3}c^{3}}\varepsilon_{F}^{3},
\label{nmaxdef}
\end{equation}
and number density in ultrarelativistic limit for thermal state is
\begin{equation}
n_{th}=\frac{12\pi\zeta(3)}{h^{3}c^{3}}(kT)^{3},
\label{nth}
\end{equation}
where $\varepsilon_{F}$ is the Fermi energy, which plays a role of an upper
particle energy boundary.

\begin{figure}[ptb]
\begin{center}
\includegraphics[width=80mm]{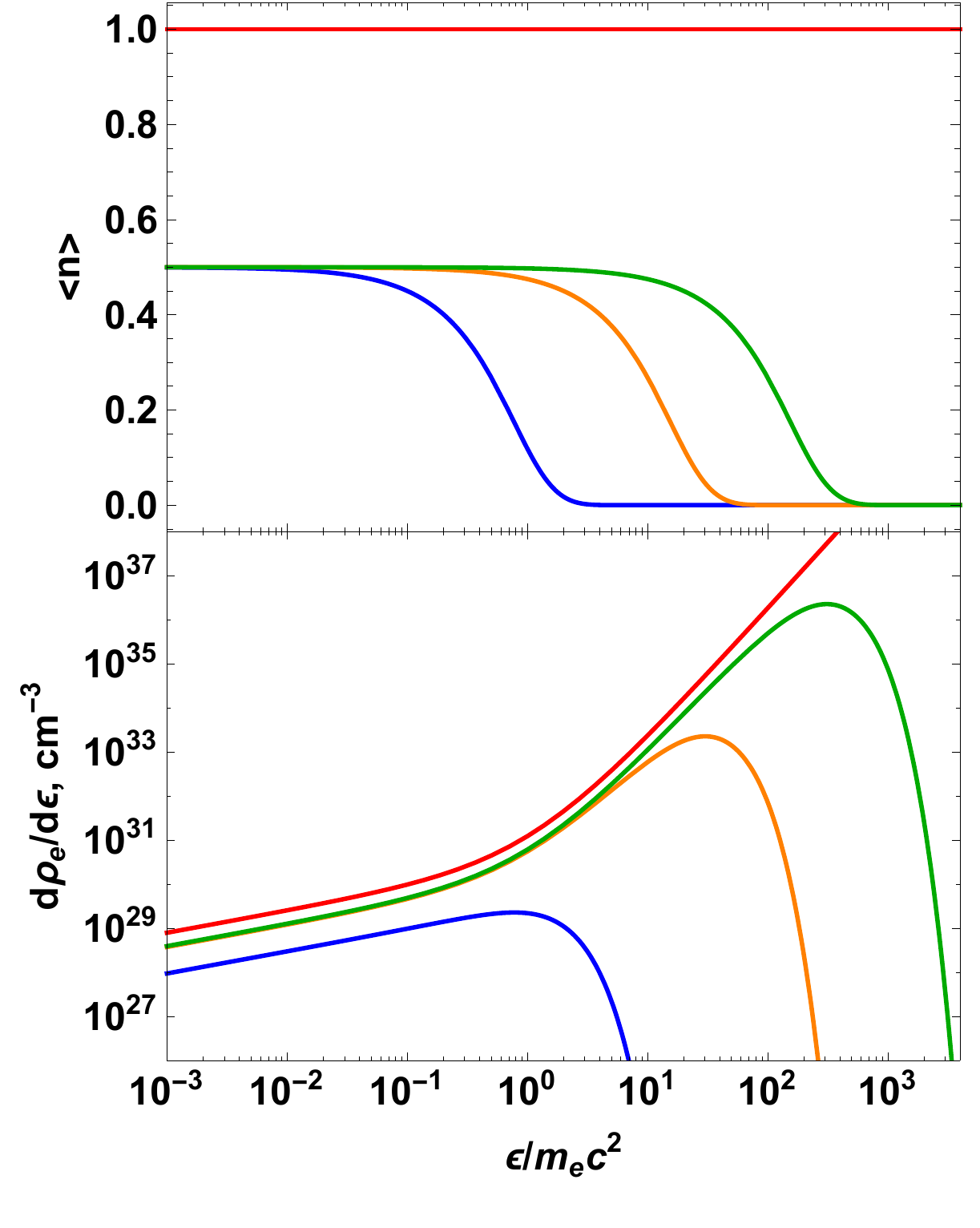}
\end{center}
\caption{Thermal average  occupation numbers (top) and thermal spectral energy
density (bottom) of pairs for selected temperatures: $\theta=0.5$ (blue), $\theta=10$ (orange), $\theta=100$ (green). The limiting spectral density for pairs according to Pauli principle is shown in red.}%
\label{occupspe}
\end{figure}

Occupation number of pairs $<n>=h^3f/g$, where $f$ is distribution function, $g$ is the number of helicity states is shown in Fig. \ref{occupspe} (top) for selected temperatures, along with the corresponding spectral energy density $d\rho/d\epsilon$ , see Fig. \ref{occupspe} (bottom). It is clear that for $\varepsilon\ll kT$ in thermal state $\langle
n\rangle\simeq1/2$.

\begin{figure}[ptb]
\begin{center}
\includegraphics[width=80mm]{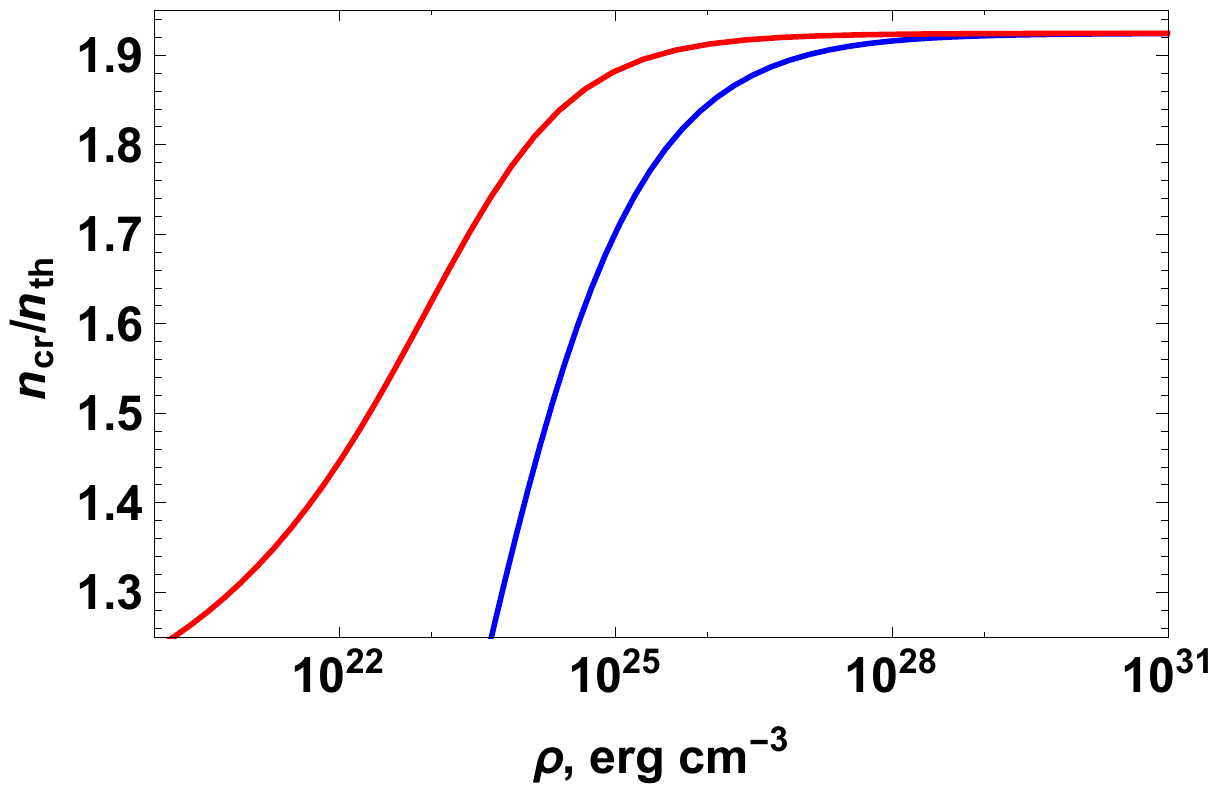}
\end{center}
\caption{The ratio $n_{cr}/n_{th}$ computed numerically (red) and analytically in ultrarelativistic approximation (blue).}%
\label{cncrit}%
\end{figure}

To compare $n_{cr}$ with $n_{th}$ we solve equation $\rho_{cr}=\rho_{th}$
for $\varepsilon_{F}$, so we obtain $n_{cr}$ as a function of total energy
density $\rho$ (or final equilibrium temperature $\theta$). On Fig. \ref{cncrit} we show the ratio $n_{cr}/n_{th}$ as computed from the integral over the distribution function (red) and as computed from ultrarelativistic expressions, eq. (\ref{nth}) and (\ref{nmaxdef}) (blue).

The relation $n_{cr}/n_{th}$ is limited with high energy
asymptotic $\frac{2^{1/4}7^{3/4}\pi^{3}}{9\times15^{3/4}\zeta(3)}\simeq1.9$.
The last result shows that photonless plasma state cannot have $D\ll1$. In other words, highly
degenerate plasma should contain large number of photons in addition to pairs.

\section{Relativistic Boltzmann equations}

In order to study the influence of quantum degeneracy on thermalization process of electron-positron-photon plasma one should solve relativistic Boltzmann equations with Pauli blocking and Bose enhancement factors in collision integrals (Uehling-Uhlenbeck equations). In this work we focus on highly degenerate pair state and the effect of the Pauli blocking on plasma thermalization. Time evolution of one-particle distribution functions of electrons $e^{-}$, positrons $e^{+}$
and photons $\gamma$ are found by numerical integration of relativistic
Boltzmann equations \cite{bookveresh} including quantum corrections
\begin{equation}
\frac{1}{c}\frac{\partial f_{i}}{\partial t}=\sum_{q}\left(  \eta_{i}^{q}%
-\chi_{i}^{q}f_{i}\right) ,\label{BE}%
\end{equation}
where $f_{i}(\epsilon,t)$ are their distribution functions, index $i$ denotes
the sort of particles, $\epsilon$ is their energy, $\eta_{i}^{q}$ and
$\chi_{i}^{q}$ are the emission and the absorption coefficients of a particle
of type "$i$" via the physical process labelled by $q$, $c$ is the speed of
light. The emission and absorption coefficients for the particle $I$ in a
binary process $I+II\leftrightarrows III+IV$ have the following form:
\begin{align}
\label{eta2p}
\eta_{I}^{2p}  &  =  \int d^{3}p_{2} d^{3}p_{3} d^{3}p_{4}
\ W_{(3,4|1,2)} \ f_{III} f_{IV} \\ & \times \left( 1+\xi f_{I}\right)  \left( 1+\xi
f_{II}\right) ,\nonumber
\end{align}
\begin{align}
\label{chi2p}
\chi_{I}^{2p}f_{I}  &  =  \int d^{3}p_{2} d^{3}p_{3} d^{3}p_{4}
\ W_{(1,2|3,4)} \ f_{I} f_{II} \\ & \times \left( 1+\xi f_{III}\right)  \left( 1+\xi
f_{IV}\right) ,\nonumber
\end{align}
where transition rates are $W_{(3,4|1,2)}d^{3}p_{3}d^{3}p_{4}=V dw_{(3,4|1,2)}%
$ and $W_{(1,2|3,4)}d^{3}p_{1}d^{3}p_{2}=V dw_{(1,2|3,4)}$, $V$ is
normalization volume, $dw $ is differential reaction probability per unit
time, $\xi=\psi h^{3}/2$ and $\psi$ is +1,-1,0 for Bose-Einstein, Fermi-Dirac,
Maxwell-Boltzmann statistic, respectively. In what follows we refer to these
cases as quantum ($\psi=\pm1$) and classical ($\psi=0$), respectively, $h$ is
Planck's constant.

The emission and absorption coefficients for the particle $I$ in a triple
process $I+II\leftrightarrows III+IV+V$ have the following form:
\begin{align}
\label{eta3p}
\eta_{I}^{3p} & = \int d^{3}p_{2} d^{3}p_{3} d^{3}p_{4} d^{3}p_{5}
\ W_{(3,4,5|1,2)} \ f_{III} f_{IV} f_{V} \\ & \times \left(  1+\xi f_{I}\right)  \left(
1+\xi f_{II}\right) ,\nonumber
\end{align}
\begin{align}
\label{chi3p}
\chi_{I}^{3p}f_{I} & = \int d^{3}p_{2} d^{3}p_{3} d^{3}p_{4} d^{3}p_{5}
\ W_{(1,2|3,4,5)} \\ & \times f_{I}f_{II}\left( 1+\xi f_{III} \right) \left( 1+\xi
f_{IV} \right) \left( 1+\xi f_{V} \right) ,\nonumber
\end{align}
where $W_{(3,4,5|1,2)}d^{3}p_{3}d^{3}p_{4}d^{3}p_{5}=V dw_{(3,4,5|1,2)}$ and
$W_{(1,2|3,4,5)}d^{3}p_{1}d^{3}p_{2}=V^{2} dw_{(1,2|3,4,5)}$. The expression
for $dw$ is given in QED as:
\begin{align}
dw & =  c(2\pi\hbar)^{4} \delta(\epsilon_{in}-\epsilon_{f\!in})\delta
(\mathbf{p}_{in}-\mathbf{p}_{f\!in})|M_{fi}|^{2} V \\ & \times\left(  \prod_{in}
\frac{\hbar c}{2\epsilon_{in} V}\right)  \left(  \prod_{f\!in} \frac
{d^{3}p_{f\!in}}{(2\pi\hbar)^{3}} \frac{\hbar c}{2\epsilon_{f\!in}}\right) ,\nonumber
\end{align}
where $\mathbf{p}_{f\!in}$ and $\epsilon_{f\!in}$ are respectively momenta and
energies of outgoing particles, $\mathbf{p}_{in}$ and $\epsilon_{in}$ are
momenta and energies of incoming particles, $M_{fi}$ is the corresponding
matrix element, $\delta$-functions stand for energy-momentum conservation,
$\hbar=h/2\pi$. Therefore, collision integrals, i.e. right-hand side of eqs.
(\ref{BE}), are integrals over the phase space of interacting particles, which
include the QED matrix elements, see e.g. \cite{Berestetskii1982,bookveresh}
for binary reactions and \cite{1952RSPSA.215..497M} for double Compton
scattering, \cite{2004epb..book.....H} for relativistic bremsstrahlung and
\cite{1976spr..book.....J} for substitution rules in computation of remaining
matrix elements for triple reactions. We consider all binary and triple
interactions between electrons, positrons and photons as listed in Tab.
\ref{tab1}.
\begin{table}[ptb]
\centering
\begin{tabular}
[c]{|c|c|}\hline
Binary processes & Triple processes\\\hline\hline
{M{\o }ller, Bhabha} & {Bremsstrahlung}\\
{$e^{\pm}{e^{\pm\prime}\leftrightarrow e^{\pm}}^{\prime\prime}$}${e^{\pm}%
}^{\prime\prime\prime}$ & {$e^{\pm}e^{\pm\prime}{\leftrightarrow}e^{\pm
\prime\prime}e^{\pm\prime\prime\prime}\gamma$}\\
{$e^{\pm}{e^{\mp}\leftrightarrow e^{\pm\prime}}$}${e^{\mp}}^{\prime}$ &
{$e^{\pm}e^{\mp}{\leftrightarrow}e^{\pm\prime}e{^{\mp\prime}}\gamma$}\\\hline
Single {Compton} & {Double Compton}\\
{\ $e^{\pm}\gamma{\leftrightarrow}e^{\pm}\gamma^{\prime}$} & {$e^{\pm}%
\gamma{\leftrightarrow}e^{\pm\prime}\gamma^{\prime}\gamma^{\prime\prime}$%
}\\\hline
{Pair production} & Radiative pair production,\\
and annihilation & triplet production\\
{} & and three photon annihilation\\
{$\gamma\gamma^{\prime}{\leftrightarrow}e^{\pm}e^{\mp}$} & $\gamma
\gamma^{\prime}${${\leftrightarrow}e^{\pm}e^{\mp}$}$\gamma^{\prime\prime}$\\
& $e^{\pm}\gamma${${\leftrightarrow}e^{\pm\prime}{e^{\mp}}e^{\pm\prime\prime}
$}\\
& {$e^{\pm}e^{\mp}{\leftrightarrow}\gamma\gamma^{\prime}$}$\gamma
^{\prime\prime}$\\\hline
\end{tabular}
\caption{Binary and triple QED processes in the pair plasma.}%
\label{tab1}%
\end{table}

Details of numerical integration scheme which solves the coupled system of integro-differential equations (\ref{BE}) on the grid in the phase space using a finite difference method is presented in \cite{2018JCoPh.373..533P}.

\section{Thermalization of superdegenerate plasma}

Thermalization process with quantum degeneracy both in non-relativistic and relativistic cases was investigated in \cite{pla2019}.
There the relative role of binary and triple interactions has been studied. The only case with strongly degenerate conditions considered in that work was the case with the dominance of photons both in energy and in number. 
In this work we are interested in thermalization process of pair plasma with
initially strongly degenerate distribution of electron-positron pairs. In
addition, we require that initial conditions correspond to $D\ll1$. Based on
the results of the previous section, the contribution of pairs correspond to
red dots in Fig. \ref{initcond}. It is clear that the density of photons in such a
state should be larger than the density of pairs, and energy density of
photons should be less than energy of pairs. The latter condition arises because
the contribution of photons to the total energy density should be minimized,
while their density should be maximized. In what follows we analyze specific
initial conditions, represented in Fig. \ref{initcond} by blue and red dots. We shall
call this state a \emph{superdegenerate} state.

With the goal to have $D\ll1$ in the initial state we choose initial photon distribution as a delta function at the lowest energy
state on our finite numerical grid. The distribution of pairs is chosen to be at equilibrium with
zero temperature and non-zero chemical potential equal to the Fermi energy. We
note that total critical pair number can be obtained with some nonequilibrium
spectrum, which does not represent a degenerate state. Such cases are
presented on figures below by solid and dashed curves.

Below we discuss four characteristic cases of initial conditions indicated by blue dots in the
number density - energy density diagram in Fig.~\ref{initcond}:
nonrelativistic and relativistic plasma with $D>1$ and $D<1$. It is appropriate to differentiate 
nonrelativistic and relativistic plasma by means of its average particle energy
in thermal equilibrium: in nonrelativistic case this quantity is less than electron rest energy
($\rho_{tot}^{th}< m_{e} c^{2}n_{tot}^{th}$), and in relativistic case otherwise.

\begin{figure}[ptb]
\begin{center}
\includegraphics[width=80mm]{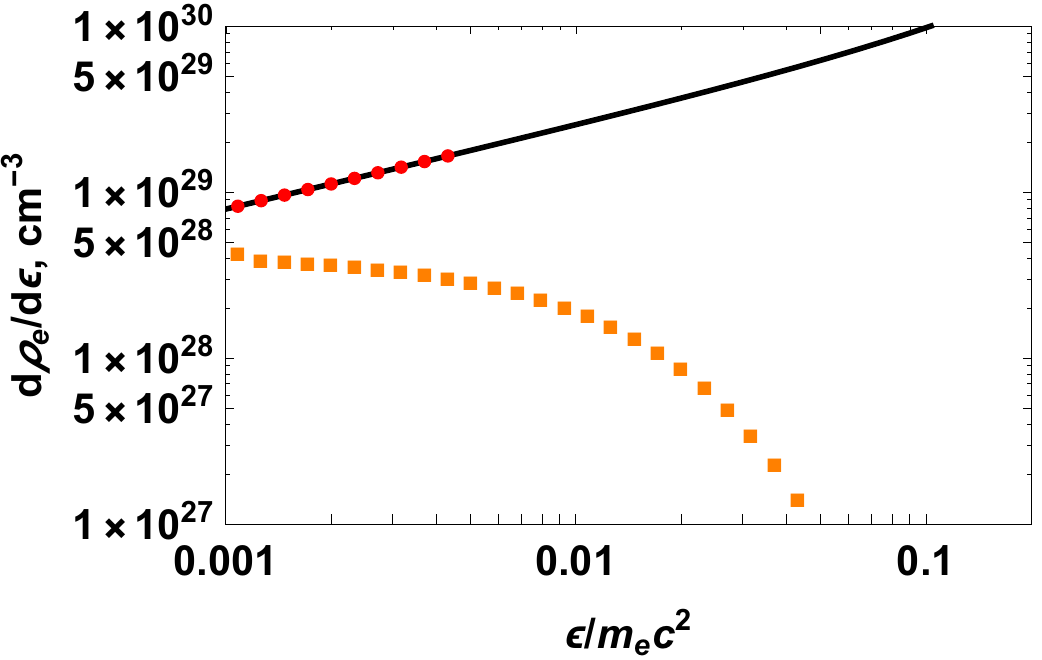} .
\end{center}
\caption{Initial electron/positron spectral energy density: critical
energy density with $\varepsilon_{F}=0.0043 m_{e} c^{2}$ (red), power law
spectral energy density $d\rho/d\varepsilon=a(\varepsilon/\varepsilon_{0}%
)^{b}, \text{with}~a=2.778~\text{cm}^{-3}$ and $b=-81,~\varepsilon
_{0}=10^{-5.74}~\text{erg}$ (orange). Horizontal axis shows kinetic energy without
electron rest energy $m_{e} c^{2}$.}%
\label{initspectrae}
\end{figure}
\begin{figure}[ptb]
\begin{center}
\includegraphics[width=80mm]{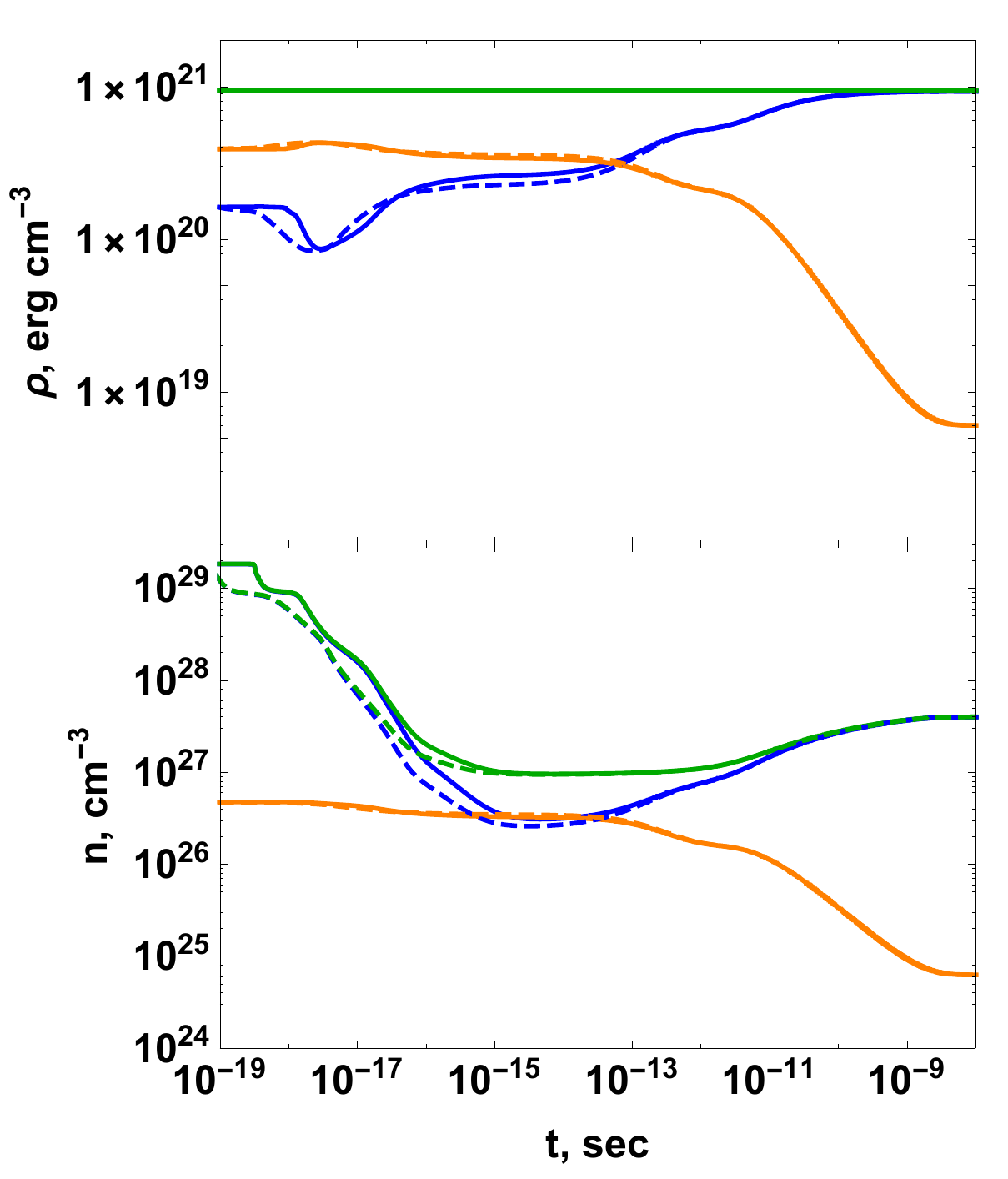}
\end{center}
\caption{Time evolution of energy density (top) and particle number density (bottom) for
nonrelativistic pair plasma with degenerate initial pair state (solid) and
nondegenerate initial pair state (dashed). Final equilibrium temperature is
$\theta_{f\!in}=0.1$.}%
\label{rhon_nonrel_pair}%
\end{figure}

\begin{figure}[ptb]
\begin{center}
\includegraphics[width=80mm]{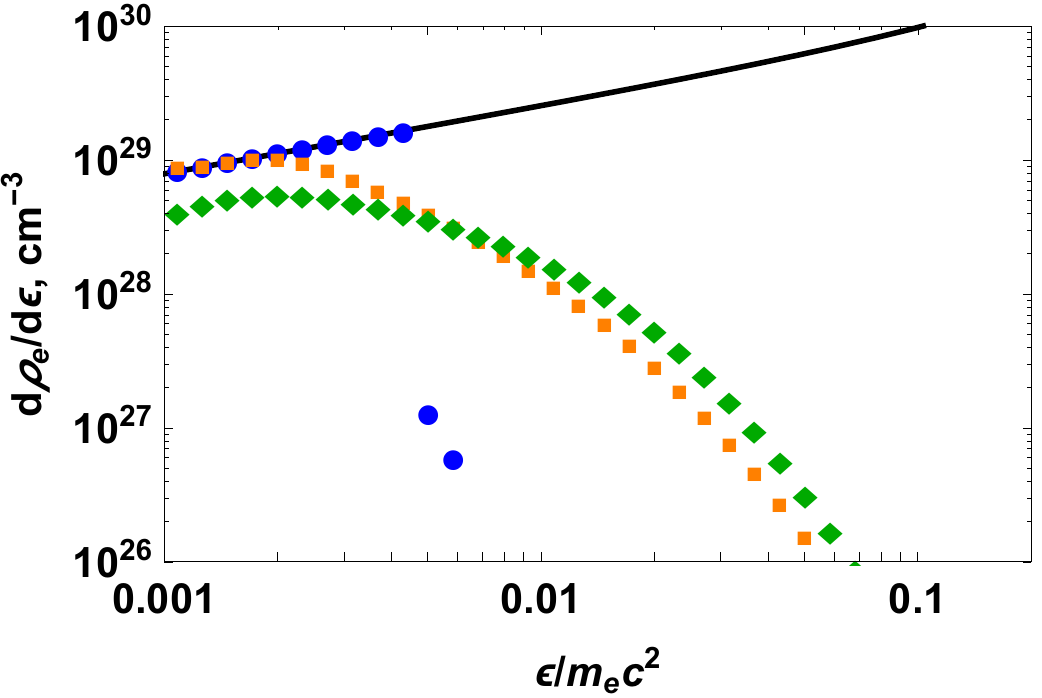}
\end{center}
\caption{Time evolution of electron/positron spectral energy density for
nonrelativistic pair plasma with degenerate initial pair state for selected
time moments: $t=1\times10^{-19}~\text{sec}$ (blue), $t=8\times
10^{-19}~\text{sec}$ (orange) and $t=9\times10^{-19}~\text{sec}$ (green). Final equilibrium
temperature is $\theta_{f\!in}=0.1$. Horizontal axis shows kinetic energy
without electron rest energy $m_{e} c^{2}$.}%
\label{spectraeoft}%
\end{figure}

First, we show the result of the simulation for non-relativistic pair plasma
with total energy density $\rho_{tot}=9.4\times10^{20}\,\text{erg cm}^{-3}$
corresponding to a final equilibrium temperature $\theta_{f\!in}=0.1$. Total
initial particle number density is $n_{tot}^{in}=45 n_{tot}^{f\!in}$, where
$n_{tot}^{f\!in}=4.2\times10^{27}\,\text{cm}^{-3}$ is the final total particle
number density in equilibrium. Two different initial spectral distribution of
pairs are considered, see Fig.~\ref{initspectrae}: a power law pair
spectrum (orange), which is far from degenerate spectrum, or a fully degenerate state (red). The time evolution of basic
thermodynamic quantities is shown on Fig.~\ref{rhon_nonrel_pair}. Solid curves
correspond to initial full degenerate pairs with zero temperature, while dashed
curves correspond to non-equilibrium non-degenerate distribution of pairs. Total energy density does not change in time
due to energy conservation. Total particle number density changes only due to
imbalance in triple processes. The kinetic equilibrium is established at
$t\simeq2\times10^{-11}$ sec. The thermal equilibrium is reached with zero
chemical potential and final temperature $\theta_{f\!in}=0.1$ at
$t\simeq10^{-8}$ sec. Spectral evolution of electrons is shown in Fig.
\ref{spectraeoft}. Note that pair annihilation is not subject to Pauli
blocking so pair annihilation leads to disappearance of degeneracy. This
result shows that thermalization process started from superdegenerate state is
not influenced by quantum corrections and it is in a full agreement with the
previous results obtained for the case of nondegenerate initial state
\cite{pla2019}. Thermalization dynamics in the case of non-equilibrium and
non-degenerate distribution of pairs with the same particle energy density and
particle number is presented by dashed curves in Fig. ~\ref{rhon_nonrel_pair}.
Both solid and dashed curves show similar evolution.

\begin{figure}[ptb]
\begin{center}
\includegraphics[width=80mm]{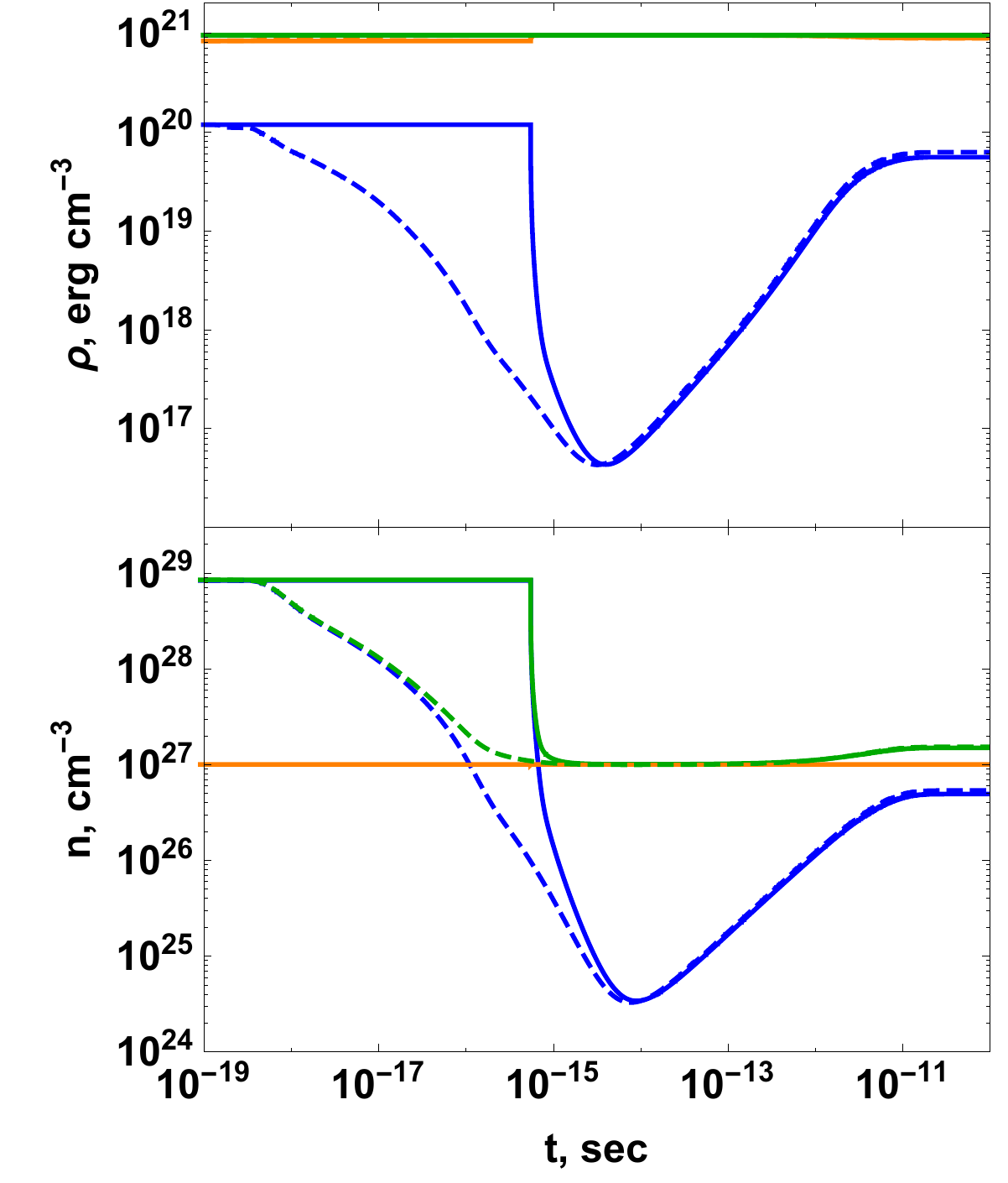}
\end{center}
\caption{Time evolution of energy density (top) and particle number density (bottom) for
nonrelativistic photon-electron plasma with degenerate initial pair state
(solid) and nondegenerate initial electron state (dashed). Final equilibrium
temperature is $\theta_{f\!in}=0.05$.}%
\label{rhon_nonrel_pairless}%
\end{figure}

Second, we turn to the case of electron-photon plasma. The positive charge, needed to compensate for electron charge, is assumed to be present, but its careers are not considered. The probability of
creation of electron-positron pairs from photons at nonrelativistic temperature
is exponentially suppressed. Then an initial fully degenerate electron state
can be preserved for a time larger compared to the characteristic pair
annihilation time. As a result, Pauli blocking multipliers might become
important for thermalization process. In Fig.~\ref{rhon_nonrel_pairless} we
show the result of the simulation for the case of nonrelativistic
photon-electron plasma with superdegenerate initial state (solid curves) and
analogous simulation with nondegenerate initial electron state (dashed
curves). Electron number is constant and photon number is changing due to
imbalance in Double Compton scattering and Bremsstrahlung processes. There is
a sharp difference between degenerate (solid curve) and non-degenerate (dashed
curve) cases. For fully degenerate initial conditions Pauli blocking
significantly reduces reaction rates (see Apendix). As a consequence kinetic
evolution starts much later, only at $t\sim10^{-15}$ sec with decrease of
thermodynamic quantities. Then electron distribution quickly establishes the
Fermi-Dirac form due to fast Coulomb scattering process. We note that photon
state at that moment is not described yet by the Bose-Einstein distribution,
because Compton scattering timescale is longer than Coulomb scattering one.
The simulation shows that balance in Compton scattering process sets at the
time moment $t\sim6 \times10^{-12}$ sec and thermal equilibrium sets at
$t\sim10^{-11}$ sec. Thus, both binary and triple processes are unbalanced
until final thermal equilibrium. 

\begin{figure}[ptb]
\begin{center}
\includegraphics[width=80mm]{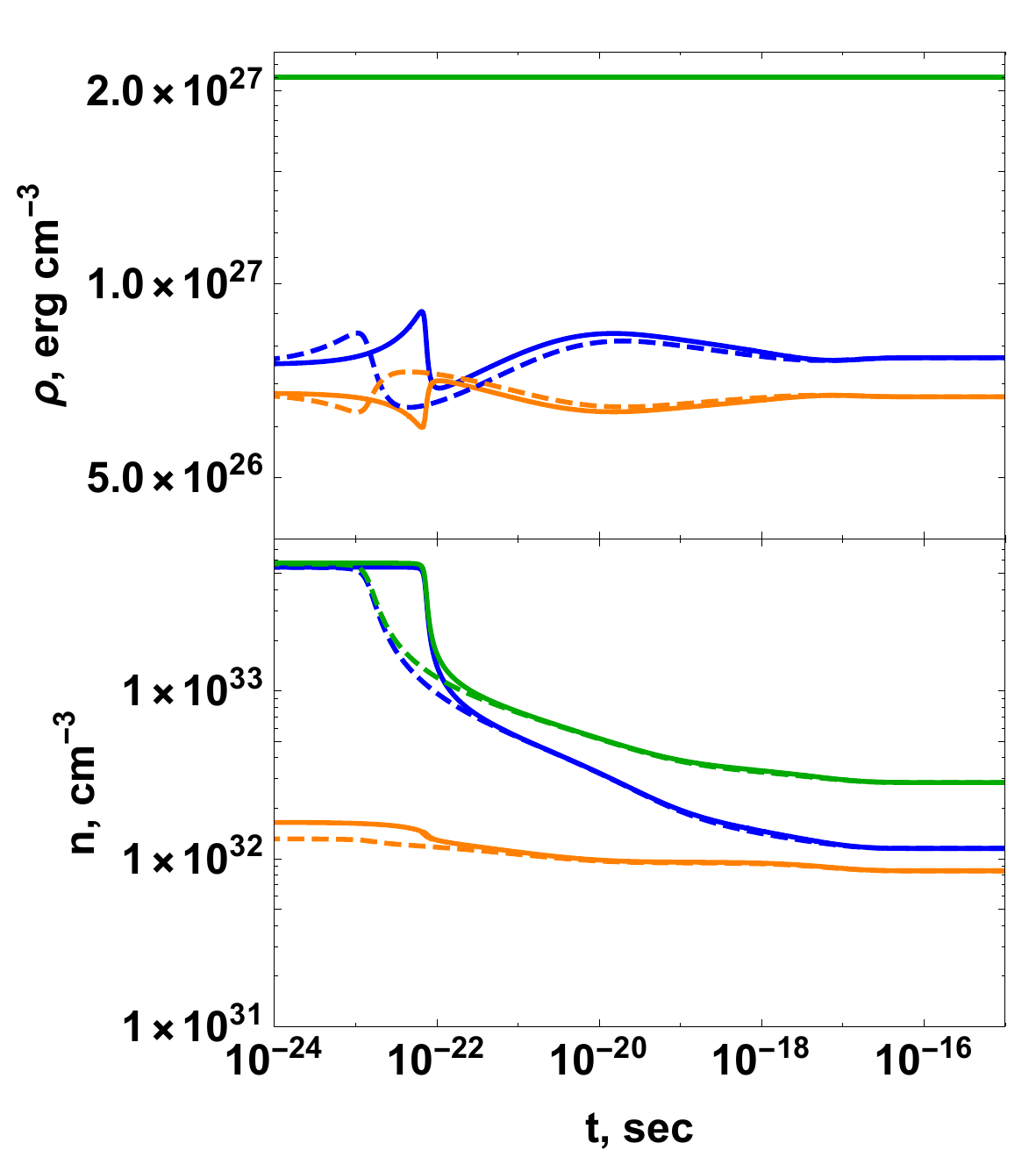}
\end{center}
\caption{Time evolution of energy density (top) and particle number density (bottom) for relativistic
pair plasma with degenerate initial pair state (solid) and nondegenerate
initial pair state (dashed). Final equilibrium temperature is $\theta
_{f\!in}=3$.}%
\label{rhon_rel_pair}%
\end{figure}

Third, we show the result of the simulation for relativistic pair plasma with
total energy density $\rho_{tot}=2.1\times10^{27}\,\text{erg cm}^{-3}$
corresponding to a final equilibrium temperature $\theta_{f\!in}=3$. Total
initial particle number density is $n_{tot}^{in}=20 n_{tot}^{f\!in}$, where
$n_{tot}^{f\!in}=2.8\times10^{32}\,\text{cm}^{-3}$ is the final total particle
number density in equilibrium. Time evolution of basic thermodynamic
quantities is shown in Fig.~\ref{rhon_rel_pair}. The kinetic equilibrium is
absent in this case. The thermal equilibrium is reached at $t\sim10^{-16}$ sec with zero chemical
potential and final temperature $\theta_{f\!in}=3$. As
in the nonrelativistic case degenerate (solid curve) and non-degenerate
(dashed curve) initial conditions lead to the similar thermalization process
due to pair annihilation process. Thus, for the case of relativistic pair
plasma thermalization process started from superdegenerate state is not
influenced by Pauli blocking. Thermalization process goes in the same way as
in the case of degenerate pairless initial state reported in \cite{pla2019}.
Note that the faster onset of the evolution of the total particle number density of particles
caused by the to Bose enhancement of two-photon annihilation rate in radiative
pair production process, see Appendix, found previously is also valid for the case of
superdegenerate initial conditions.

\begin{figure}[ptb]
\begin{center}
\includegraphics[width=80mm]{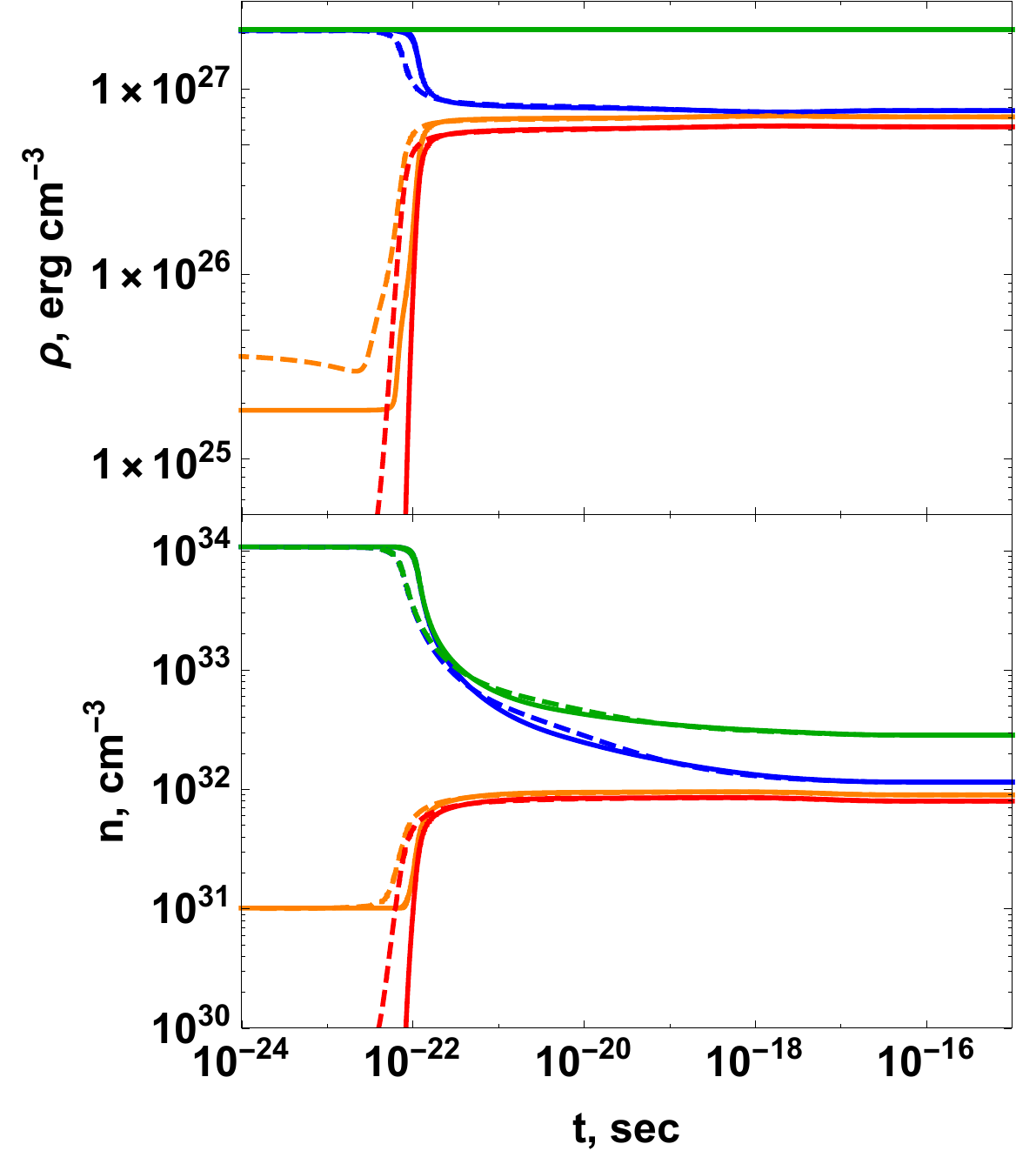}
\end{center}
\caption{Time evolution of energy density (top) and particle number density (bottom) for relativistic
transitional electron-photon plasma with degenerate initial pair state (solid)
and nondegenerate initial pair state (dashed). Final equilibrium temperature
is $\theta_{f\!in}=3$.}%
\label{rhon_rel_pairless_trans}%
\end{figure}

\begin{figure}[ptb]
\begin{center}
\includegraphics[width=80mm]{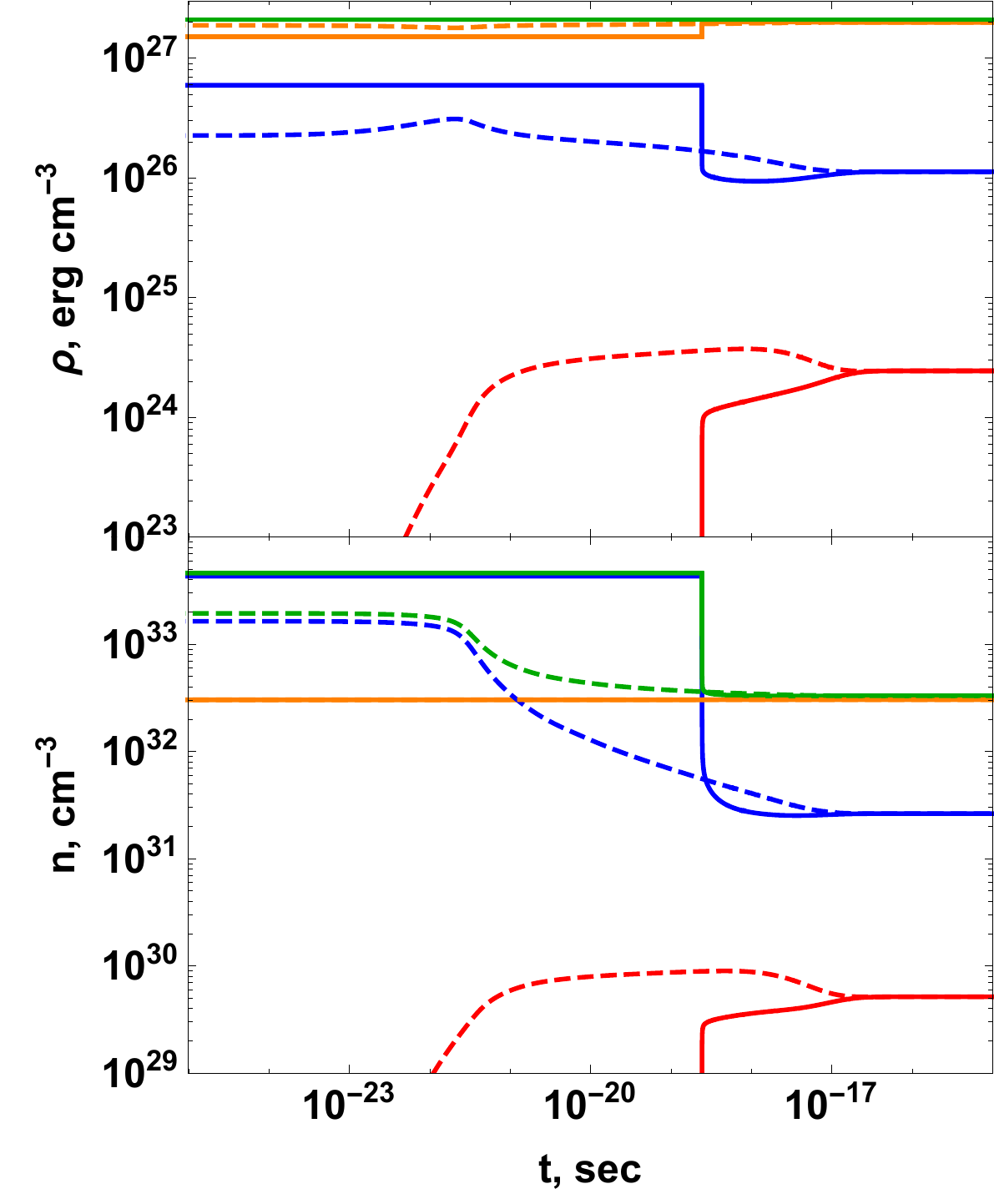}
\end{center}
\caption{Time evolution of energy density (top) and particle number density (bottom) for relativistic
photon-electron plasma with degenerate initial pair state (solid) and
nondegenerate initial electron state (dashed). Final equilibrium temperature
is $\theta_{f\!in}=1.9$.}%
\label{rhon_rel_pairless}%
\end{figure}

Finally, we discuss thermalization process with three different
cases of relativistic photon-electron plasma. In contrast to nonrelativistic
electron-photon plasma, relativistic plasma contains positrons. For a given
total energy density the number of final positrons depends on the number of
initial electrons. In equilibrium chemical potentials of electrons and
positrons are equal on magnitude and their sum is zero $(\mu_{-} + \mu_{+}%
=0)$. If for the final state $\mu_{-}>1$ then final electron number is much
greater than final positron number and plasma can be treated as
electron-photon plasma. If for the final state $\mu_{-}\ll1$ then final
electron number is approximately equal to final positron number and plasma
represents a transitional case between electron-photon plasma (when $\mu
_{-}=-\mu_{+}>1$) and pair plasma (when $\mu_{+}=\mu_{-}=0$). First, the
simulation for the relativistic transitional electron-photon plasma with fully
degenerate initial electrons and $D\gg1$ showed that thermalization for both
degenerate and nondegenerate initial electrons proceeds in the similar way and
Pauli blocking effects are negligible (see solid and dashed curves on
Fig.~\ref{rhon_rel_pairless_trans}). The reason is that initial electron number
is much less then final electron number ($n_{e}^{in}\ll n_{e}^{f\!in}$) and
the degree of electron degeneracy is not sufficient to block any reaction. This
initial state is not a superdegenarate state because of violation of the
condition $\rho_{e}^{in}>\rho_{\gamma}^{in}$. Second, for the initial state with
a large number of degenerate electrons (when $n_{e}^{in}\simeq n_{e}^{f\!in}$)
and with highly energetic photons (when $D\gtrsim1$) the Pauli
blocking effects are also negligible, because degeneracy of electrons vanishes due to Compton
scattering (energetic photons scatter electrons to high energies). Third, 
we selected initial state containing a large number of degenerate electrons (when
$n_{e}^{in}\simeq n_{e}^{f\!in}$) and low energetic photons, that is the
case $D\ll1$ and present thermalization dynamics in Fig. ~\ref{rhon_rel_pairless}. As energy of initial photons is much less than $m c^{2}$, pair creation process is suppressed and initial degenerate electron state can
be preserved until photons acquire energy more than $m c^{2}$. For the
simulation the following initial conditions are chosen: total energy density is
$\rho_{tot}=2.1\times10^{27}\,\text{erg cm}^{-3}$ corresponding to a final
equilibrium temperature $\theta_{f\!in}=1.9$, total initial particle number
density is $n_{tot}^{in}=40 n_{tot}^{f\!in}$, where $n_{tot}^{f\!in}%
=3.2\times10^{32}\,\text{cm}^{-3}$. One can see a sharp decrease of
thermodynamic quantities at the time moment $t\sim2\times10^{-19}$ sec,
degenerate electron spectrum is preserved until the same time. The creation of
positrons starts at the same moment. Thermalization process has an avalanche-like
character. At the same time, simulation with nondegenerate initial electrons shows a smooth
monotonic thermalization process (dashed curves in Fig.~\ref{rhon_rel_pairless}) which starts much earlier (solid curves on Fig.~\ref{rhon_rel_pairless}).

\section{Summary}

In this work the influence of Pauli blocking on thermalization process of
relativistic plasma is studied for the first time using relativistic
Uehling-Uhlenbeck equations taking into account all binary and triple QED
processes. A wide range of initial conditions is analyzed, and several characteristic cases are presented. Both
electron-positron-photon plasma, and photon-electron
plasma are studied in relativistic and non relativistic domains.

We show that fully degenerate electron-positron state without photons
corresponds to the degeneracy parameter $D\gtrsim1$. Thus, in order study plasma evolution with $D\ll1$ initial state should contain photons, in addition to degenerate electrons and positrons. The energy density and particle number
density of plasma components in a such a superdegenerate state should satisfy
the conditions: $\rho_{\pm}\gg\rho_{\gamma}$ and $n_{\pm}\ll n_{\gamma}$.

We found that thermalization process of superdegenerate pair plasma
with total energy density ranging from $10^{20}~\text{erg cm}^{-3}$ to
$10^{28}~\text{erg cm}^{-3}$ is not influenced by Pauli blocking and
relaxation proceeds similarly both for $D\ll1$ and for $D\gtrsim1$. Pair annihilation process, not affected by the Pauli blocking, plays a crucial role here. This process quickly destroys initial degeneracy
of electrons and positrons diminishing the role of the Pauli blocking effect.

At the same time, the process of thermalization in electron-photon plasma is very different. In nonrelativistic superdegenarate electron-photon plasma thermalization is delayed due to Pauli blocking of interactions with electrons. Once reactions are activated they have an avalanche-like behaviour. This is because initial state is preserved
until photons can scatter degenerate electrons above their Fermi energy, where
they states are not degenerate any more. The characteristic timescale for the
beginning of the avalanche can be estimated as inverse reaction rate of the
Compton scattering and it equals to $\tau_{aval}\simeq10^{-15}$ sec
for the number density $n_{tot}=8.5\times 10^{28}~\text{cm}^{-3}$. This is an interesting effect and it could manifest in laboratory and in astrophysical environments.

Relativistic electron-photon plasma contains positrons and initial conditions can be separated in
two different classes defined by the conditions $\mu_{e}^{f\!in}\ll1$ and
$\mu_{e}^{f\!in}>1$. In the first case final positron number density
approximately equals to final electron number density. At the same time the
number of initial degenerate electrons is not sufficiently large to satisfy
the condition $\rho_{e}>\rho_{\gamma}$, and thermalization process of this
type of relativistic electron-photon plasma is not affected by the Pauli
blocking. For the second type of relativistic electron-photon plasma the
superdegenerate state can exist and thermalization process in that case shows
avalanche behaviour. The characteristic timescale for the beginning of
avalanche is $\tau_{aval}\simeq10^{-19}$ sec.

\begin{acknowledgments}
This work is supported within the joint BRFFR-ICRANet-2018 funding programme.
\end{acknowledgments}

\onecolumngrid

\appendix

\begin{table}[ptb]
\center
\begin{tabular}
[c]{|l|l|l|l|}\hline
& $(\gamma\gamma^{\prime}${${\leftrightarrow}e^{\pm}e^{\mp}$}$\gamma
^{\prime\prime})$ & $e^{\pm}\gamma${${\leftrightarrow}e^{\pm\prime}{e^{\mp}%
}e^{\pm\prime\prime} $} & $e^{\pm}\gamma{\leftrightarrow}e^{\pm\prime}%
\gamma^{\prime}\gamma^{\prime\prime}$\\\hline\hline
$\eta_{\gamma}$ & $f_{\gamma}f_{\gamma}(1+f_{\gamma})(1-f_{+})(1-f_{-})$ &
$f_{\pm}f_{+} f_{-} (1+f_{\gamma})(1-f_{\pm})$ & $f_{\gamma}f_{\pm}(1-f_{\pm
})(1+f_{\gamma})(1+f_{\gamma})$\\
& $f_{\gamma}f_{+}f_{-} (1+f_{\gamma})(1+f_{\gamma})$ & $f_{\pm}f_{\gamma
}f_{\gamma}(1-f_{\pm})(1+f_{\gamma})$ & \\\hline
$\chi_{\gamma}$ & $f_{\gamma}f_{+} f_{-} (1+f_{\gamma})(1+f_{\gamma})$ &
$f_{\gamma}f_{\pm}(1-f_{\pm})(1-f_{+})(1-f_{-})$ & $f_{\pm}f_{\gamma}%
f_{\gamma}(1-f_{\pm})(1+f_{\gamma})$\\
& $f_{\gamma}f_{\gamma}(1+f_{\gamma})(1-f_{+})(1-f_{-})$ &  & $f_{\gamma
}f_{\pm}(1-f_{\pm})(1+f_{\gamma})(1+f_{\gamma})$\\\hline
$\eta_{\pm}$ & $f_{\gamma}f_{\gamma}(1+f_{\gamma})(1-f_{+})(1-f_{-})$ &
$f_{\gamma}f_{\pm}(1-f_{\pm})(1-f_{+})(1-f_{-})$ & $f_{\gamma}f_{\pm}%
(1-f_{\pm})(1+f_{\gamma})(1+f_{\gamma})$\\
&  & $f_{\pm}f_{+} f_{-} (1+f_{\gamma})(1-f_{\pm})$ & $f_{\pm}f_{\gamma
}f_{\gamma}(1-f_{\pm})(1+f_{\gamma})$\\\hline
$\chi_{\pm}$ & $f_{\gamma}f_{+}f_{-}(1+f_{\gamma})(1-f_{\gamma})$ & $f_{\pm
}f_{+} f_{-} (1+f_{\gamma})(1-f_{\pm})$ & $f_{\pm}f_{\gamma}f_{\gamma
}(1-f_{\pm})(1+f_{\gamma})$\\
&  & $f_{\gamma}f_{\pm}(1-f_{\pm})(1-f_{+})(1-f_{-})$ & $f_{\gamma}f_{\pm
}(1-f_{\pm})(1+f_{\gamma})(1+f_{\gamma})$\\\hline\hline
& $e^{\pm}e^{\pm\prime}{\leftrightarrow}e^{\pm\prime\prime}e^{\pm\prime
\prime\prime}\gamma$ & {$e^{\pm}e^{\mp}{\leftrightarrow}\gamma\gamma^{\prime}
$}$\gamma^{\prime\prime}$ & \\\hline\hline
$\eta_{\gamma}$ & $f_{\pm}f_{\pm}(1-f_{\pm})(1-f_{\pm})(1+f_{\gamma})$ &
$f_{+} f_{-} (1+f_{\gamma})(1+f_{\gamma})(1+f_{\gamma})$ & \\\hline
$\chi_{\gamma}$ & $f_{\pm}f_{\pm}f_{\gamma}(1-f_{\pm})(1-f_{\pm})$ &
$f_{\gamma}f_{\gamma}f_{\gamma}(1-f_{+})(1-f_{-})$ & \\\hline
$\eta_{\pm}$ & $f_{\pm}f_{\pm}(1-f_{\pm})(1-f_{\pm})(1+f_{\gamma})$ &
$f_{\gamma}f_{\pm}(1-f_{\pm})(1-f_{+})(1-f_{-})$ & \\
&  & $f_{\pm}f_{\pm}f_{\gamma}(1-f_{\pm})(1-f_{\pm})$ & \\\hline
$\chi_{\pm}$ & $f_{\pm}f_{\pm}f_{\gamma}(1-f_{\pm})(1-f_{\pm})$ & $f_{+} f_{-}
(1+f_{\gamma})(1+f_{\gamma})(1+f_{\gamma})$ & \\
& $f_{\pm}f_{\pm}(1-f_{\pm})(1-f_{\pm})(1+f_{\gamma})$ &  & \\\hline\hline
& $e^{\pm}e^{\mp}{\leftrightarrow}\gamma\gamma^{\prime}$ & $e^{\pm}
\gamma{\leftrightarrow}e^{\pm\prime}\gamma^{\prime}$ & $e^{\pm} e^{\pm
}{\leftrightarrow}e^{\pm\prime}e^{\pm\prime}$\\\hline\hline
$\eta_{\gamma}$ & $f_{+} f_{-} (1+f_{\gamma})(1+f_{\gamma})$ & $f_{\pm
}f_{\gamma}(1-f_{\pm})(1+f_{\gamma})$ & $f_{\pm}f_{\pm}(1-f_{\pm})(1-f_{\pm}%
)$\\\hline
$\chi_{\gamma}$ & $f_{\gamma}f_{\gamma}(1-f_{+})(1-f_{-})$ & $f_{\pm}%
f_{\gamma}(1-f_{\pm})(1+f_{\gamma})$ & $f_{\pm}f_{\pm}(1-f_{\pm})(1-f_{\pm}%
)$\\\hline
$\eta_{\pm}$ & $f_{\gamma}f_{\gamma}(1-f_{+})(1-f_{-})$ & $f_{\pm}f_{\gamma
}(1-f_{\pm})(1+f_{\gamma})$ & $f_{\pm}f_{\pm}(1-f_{\pm})(1-f_{\pm})$\\\hline
$\chi_{\pm}$ & $f_{+} f_{-} (1+f_{\gamma})(1+f_{\gamma})$ & $f_{\pm}f_{\gamma
}(1-f_{\pm})(1+f_{\gamma})$ & $f_{\pm}f_{\pm}(1-f_{\pm})(1-f_{\pm})$\\\hline
\end{tabular}
\caption{Bose enhancement and Pauli blocking multipliers for collision
integral of Uehling-Uhlenbeck equation for a given QED process. Here constant
$h^{3}/2$ is omitted.}%
\label{tab2}%
\end{table}

\bibliographystyle{aip}

\end{document}